\documentclass[aps,pra,twocolumn,superscriptaddress,showpacs,floatfix]{revtex4-1}
\usepackage{times,amssymb,amsmath}
\usepackage{simplewick}
\usepackage[pdftex]{graphicx}
\usepackage{txfonts}
\renewcommand{\vec}[1]{\mathbf{#1}}
\newcommand{\be}{\begin{equation}}
\newcommand{\ee}{\end{equation}}
\newcommand{\bea}{\begin{eqnarray}}
\newcommand{\eea}{\end{eqnarray}}
\newcommand{\mC}{\mathcal{C}}
\newcommand{\plus}{ {\scriptscriptstyle +}}
\newcommand{\minus}{{\scriptscriptstyle -}}
\newcommand{\ket}{\rangle}
\newcommand{\bra}{\langle}
\newcommand{\nn} {\nonumber}

\begin{document}
\title{Single and double photoemission and generalizations}
\author{Y. Pavlyukh}
\email[]{yaroslav.pavlyukh@physik.uni-halle.de}
\affiliation{Institut f\"{u}r Physik, Martin-Luther-Universit\"{a}t
  Halle-Wittenberg, 06120 Halle, Germany}
\begin{abstract}
A unified diagrammatic treatment single and double electron photoemission currents is
presented. The lesser density-density response function is the starting point of these
derivations. Diagrams for higher order processes in which several electrons are observed
in coincidence can likewise be obtained. For physically relevant situations in which the
photoemission cross-section can be written as the Fermi Golden rule, the diagrams from the
nonequilibrium Green's function approach can be put direct correspondence with that of the
scattering theory.

\end{abstract}
\pacs{71.10.-w,31.15.A-,73.22.Dj}
\maketitle
\section{Introduction} 
The electron photoemission is a process in which classical or quantized electromagnetic
field interacts with a many-body target leaving the system in an unbound electronic
state~\cite{hufner_photoelectron_2003}. This subclass of optical absorption processes can
further be cathegorized according to the number of emitted particles in one scattering
event, i.e., single, double or, more generally, $n$-electron photoemission. The exact
quantum state the ionized target was left in is typically not known. However, since
detectors can determine the asymptotic scattering state of emitted electrons some
information about the ionized target can be inferred by virtue of energy and momentum
conservation laws. In particular, one can tell whether the target was left in the ground
or an excited state. Thus, we can distinguish between the no-loss photoelectron currect
and that incorporating losses. The processes leading to the emission of an electron with a
reduced energy can have various origins. On the basis of the empirical three step
model~\cite{berglund_photoemission_1964} we have a picture where a liberated electron
propagates through the sample loosing its energy on the way. In this way the
electron-photon interaction and the inelastic electron scattering are considered as two
independent processes and for the latter one speaks about the \emph{extrinsic} energy
losses~\cite{chang_deep-hole_1972,chang_deep-hole_1973}. Besides electronic excitations
the phonon~\cite{kane_simple_1966}, impurity scattering~\cite{chulkov_electronic_2006} and
Auger processes~\cite{verdozzi_correlation_1997,verdozzi_auger_2001} are important.

 On the other hand electron-electron interaction in the system causes the spectral
 function (or the density of states) to deviate from the non-interacting one in which all
 spectral weight is confined to a quasiparticle peak. A typical picture would be a
 quasiparticle peak followed by a sequence of satellite peaks. Their origin can be
 different as
 well~\cite{aryasetiawan_multiple_1996,pavlyukh_time_2013,story_cumulant_2014}. For
 instance in the photoemission from a deep core level it is the density oscillations in
 the system\,---\,plasmons\,---\,that form scattering channels responsible for the
 occurrence of plasmonic satellites.  $S$-model proposed by
 Lundqvist~\cite{lundqvist_characteristic_1969} and solved by
 Langreth~\cite{langreth_singularities_1970} captures exactly this effect and predicts a
 sequence equally spaced satellite peaks with oscillator strengths satisfying the Poisson
 distribution.  If an electron originates from one of such peaks it will come to a
 detector with a reduced energy. One speaks here about the \emph{intrinsic} energy losses
 because it is the ground state spectral density that is modified by the plasmon
 scattering~\cite{penn_theory_1977,penn_role_1978}.

Although first microscopic theories of
photoemission~\cite{schaich_model_1971,mahan_theory_1970} have taken underlying electronic
structure into account they have not provided a description of inelastic energy losses
experienced by the liberated electron. Such formulation was achieved with the help of
non-equilibrium Green's function approach by Caroli \emph{et
  al.}~\cite{caroli_inelastic_1973}. The starting point of their treatment is the zeroth
order triangular diagram where two vertices are situated on the forward and backward
tracks of the Keldysh contour~\cite{stefanucci_nonequilibrium_2013} and describe the
light-matter interaction, whereas the third vertex lying at the common point of the two
tracks in the remote future describes the detector state. There are no interaction lines
because the photoemission is possible even for non-interacting systems. Each
electron-photon interaction vertex is associated with the mimimal coupling part of the
Hamiltonian and is linearly proportional to the field amplitude.  The whole diagram, as
expected, is quadratic in the field strength and is linearly proportional to the intensity
or to the number of absorbed photons.

Progressing diagrammatically further~\cite{fujikawa_many-body_2002} one realizes that
dressing of this diagram leads not only to the renormalization of each electronic
propagator involved, but also generates vertex functions~\cite{almbladh_theory_1985}. All
these ingredients are important in different physical scenarios, they account for the
effects of optical field
screening~\cite{almbladh_importance_1986,krasovskii_dielectric_2010,uimonen_ultra-nonlocality_2014},
intrisic and extrinsic losses as well as their
interference~\cite{campbell_interference_2002,guzzo_valence_2011,guzzo_multiple_2014}, and
for the formation of the scattering states~\cite{hedin_sudden_2002}.  The goal of this
work is to give a diagrammatic description of the multiparticle emission. Before
proceeding with corresponding diagrammatic construction a short account of the single
photoemission will be given in Sec.~II. Properties of double photoemission diagrams are
discussed in Sec.~III and generalization to $n$-particle emission is presented in
Sec.~IV. The formalism looks more natural if irreducible density-density response function
describing general photoabsorption process is used as a starting point. Such point of view
has already been mentioned by Feibelman and Eastman~\cite{feibelman_photoemission_1974} in
their derivation of the Fermi Golden rule (FGR) expression for photoemission starting with
the diagrammatic approach of Caroli \emph{et al}.~\cite{caroli_inelastic_1973}. Supported
by work on positive definite diagrammatic
approximations~\cite{almbladh_theory_1985,stefanucci_diagrammatic_2014,uimonen_diagrammatic_2015}
a one-to-one correspondance between the NEGF (provided they can be written in the Fermi
Golden rule form) and Goldstone diagrams in scattering
theory~\cite{kelly_many-body_1964,lindgren_atomic_1982,amusia_atomic_1990} can established
as is outlined here.

In this work atomic units are used throughout.
\section{Ingredients of SPE\label{sec:spe}}
\begin{figure}[h!]
	\centering
       \includegraphics[width=\columnwidth]{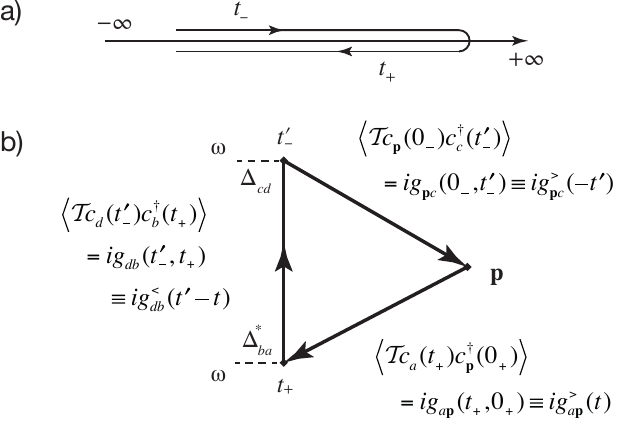}
       \caption[]{(a)The Keldysh time-loop contour $\mC$. The forward branch is denoted
         with a ``$-$'' label while the backward branch is denoted by a ``$+$'' label.
         (b) Lowest order diagram for single photoemission. $\hat\Delta$ is the operator
         for light-matter coupling associated with vertices having time arguments
         $t^\prime_{-}$ and $t_+$.}
      \label{fig:contour}
\end{figure}

 The differential cross-section fully describes the photoemission experiment. It relates
 the electron current $J_\vec{k}$ to the incoming photon flux density $F$:
\[
\frac{d\sigma}{d\vec k}=\frac{J_\vec{k}}{F}
\]
The incident photon flux is obtained by dividing the incident intensity
$I=\frac{\omega^2A^2}{2\pi c}$ by the photon energy $\omega$. The current is determined by
the electronic structure of the system and by details of its coupling to the
electromagnetic field, in velocity gauge we have $\hat \Delta=\frac1c\vec{A}\cdot\hat
{\vec p}$, where $\hat {\vec p}$ is the momentum operator.  Many-body perturbation theory
yields diagrammatic expansion of the
current~\cite{schaich_model_1971,caroli_inelastic_1973,almbladh_theory_1985,fujikawa_many-body_2002,
  uimonen_ultra-nonlocality_2014,pavlyukh_single-_2015} in terms of Green's functions on
the Keldysh conltour~\cite{stefanucci_nonequilibrium_2013}. It results  in
%==================
\be 
J_\vec{k}=\lim_{\eta\rightarrow0}2\eta
\int_{-\infty}^0\!\!d(tt')e^{\eta(t+t')}(\vec{\Delta}^* \vec{Z}^{(1)}(t,t^\prime)\vec{\Delta})_\vec{k}, 
\label{eq:Jk}
\ee
%==================
with the following ground state correlator 
%==================
\be 
Z^{(1)}(t,t^\prime)=\langle\Psi_0|c_b^\dagger(t)c_a(t)c_\vec{k}^\dagger(0)c_\vec{k}(0)
c_c^\dagger(t^\prime)c_d(t^\prime)|\Psi_0\rangle, 
\label{eq:Z}
\ee
%==================
where the field operators are in the Heisenberg representation. Three parts can be seen:
operators dependent on the times $t,\,t^\prime\in(-\infty,0]$ originate from the
  light-matter couplings $\hat
  \Delta^\dagger(t)=\sum_{ab}\Delta_{ba}^*c_b^\dagger(t)c_a(t)$ and $\hat
  \Delta(t')=\sum_{cd}\Delta_{cd}c_c^\dagger(t^\prime)c_d(t^\prime)$, whereas
  $c_\vec{k}^\dagger(0)c_\vec{k}(0)$ yields the current operator. This is the reason why
  $Z^{(1)}(t,t^\prime)$ is sometimes called the \emph{three-current} correlator.  The system
  interacts with electromagnetic field in the interval of time from remote past to the
  moment $t=0$ where the observation takes place. This entails the specific time-order of
  the operators in the correlator and is the reason why the standard time-order formalism
  is insufficient for its computation (no Kubo formula for
  photoemission~\cite{mahan_theory_1970}). On the other hand the time-ordering
  in~\eqref{eq:Z} can naturally be represented on the Keldysh contour,
  Fig.~\ref{fig:contour}(a).  The correlator can be perturbatively computed by applying
  the Wick's theorem.
%---------------------------
%---------------------------

Specification of the single-particle indices for the creation and annihilation operators
in Eq.~\eqref{eq:Z} is important. The unbound (continuum) states are denoted in bold-style
to distinguish them from general indices denoted as $a,\,b,\,c,\ldots$ that also comprise
bound states $i,\,j,\,k,\ldots$. In zeroth order there is a single diagram resulting from
the following contraction
\[
\big\langle
\contraction{c_b^\dagger(t_+)}{c_a(t_+)}{}{c^\dagger_{\vec{p}}(0)}
\contraction{c_b^\dagger(t_+)c_a(t_+)c^\dagger_{\vec{p}}(0)}{c_{\vec{p}}(0)}{}{c_c^\dagger(t^\prime_-)}
\contraction[2ex]{}{c_b^\dagger(t_+)}{c_a(t_+)c^\dagger_{\vec{p}}(0)c_{\vec{p}}(0)c_c^\dagger(t^\prime_-)}{c_d(t^\prime_-)}
c_b^\dagger(t_+)c_a(t_+)c^\dagger_{\vec{p}}(0)c_{\vec{p}}(0)c_c^\dagger(t^\prime_-)c_d(t^\prime_-)
\big\rangle.
\]
Other contractions vanish because the target is bound in its ground state, i.e.  
%---------------------------
\be
c_\vec{k}|\Psi_0\rangle=0.
\label{eq:bnd}
\ee
%---------------------------
Introducing the \emph{lesser} and \emph{greater} Green's function (GF) components 
%===========================================================
%
%===========================================================
\begin{subequations}
\label{g><}
\bea 
G^{<}_{ab}(t_1,t_2)\equiv G^{\minus\plus}_{ab}(t_1,t_2) &=&  i \bra
c^\dagger_b (t_2) c_a (t_1) \ket,
\label{g<}\\
G^{>}_{ab}(t_1,t_2) \equiv G^{\plus\minus}_{ab}(t_1,t_2) &=&-i \bra c_a (t_1) c^\dagger_b (t_2)\ket.
\label{g>}
\eea
\end{subequations}
%===========================================================
%
%===========================================================
condition~\eqref{eq:bnd} can be formulated as the requirement that lesser GFs with one or
more continuum indices vanish:
\[
G^{<}_{\vec{k} a}=G^{<}_{a \vec{k}}=0.
\]
In these notations zeroth order (in Coulomb interaction $v$) correlator is given by the
product of one lesser GF describing the propagation of an added hole and two greater GFs
describing propagation of the emitted particle [see Fig.~\ref{fig:contour}(b)]:
%---------------------------
\be
Z^{(1)}(t,t')=ig_{a\vec{p}}^>(t)g_{db}^<(t'-t)g_{\vec{p}c}^>(-t') .
\ee 
%---------------------------
The constituent GFs are mean-field ones, i.\,e. they are averages over the ground state of
corresponding Hartree-Fock Hamiltonian (we assume that it is bound and fulfills a
condition similar to ~\eqref{eq:bnd}).  The vertical electron propagator, i.\,e., the one
that connects two electron-photon vertices, is of the \emph{lesser} kind $G^<\equiv
G^{-+}$ because its first and second time arguments belong to different tracks, "$-$" and
"$+$" respectively. It contains the information on the spectral density of the occupied
states. Replacing the bare with the full propagator means that mean-field spectral density
of occupied states is replaced with the exact interacting one which also incorporates the
intrinsic energy losses. Computation of the correlated spectral densities is the main goal
of electronic structure theory. The electronic lines flowing to or from the detector are
of the \emph{greater} type and, thus, are associated with unoccupied electronic
states. Their dressing describes the formation of \emph{scattering} states observed in the
detector~\cite{bardyszewski_new_1985,hedin_optical_1990,hedin_sudden_2002}.

In addition to renormalization of fermionic lines the evaluation of next orders of
Eq.~\eqref{eq:Z} leads to appearence of vertex functions. As Almbladh
explains~\cite{almbladh_theory_1985,almbladh_importance_1986} they amount to modification
of the bare light-matter coupling operator $\hat{\Delta}$ by the vector-coupling vertex
function $\hat{\Lambda}(\epsilon+\omega,\epsilon)$. All other possible renormalizations of
$Z^{(1)}(t,t')$ contain renormalized interaction lines connecting points on the forward
and backward branches of the contour and are associated with extrinsic energy losses. In
the next section we consider a particular example of such processes involving emission of
a secondary electron.
\section{Double photoemission}

\begin{figure}[t!]
	\centering
       \includegraphics[width=\columnwidth]{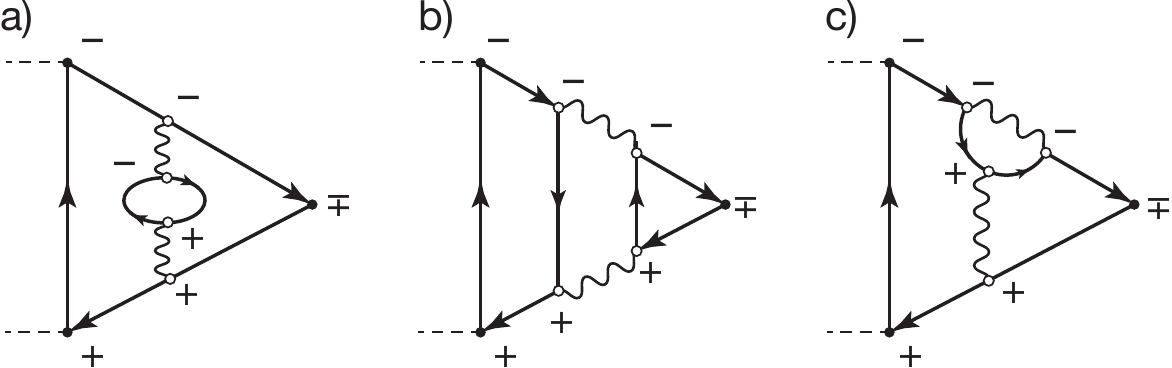}
       \caption[]{Lowest order diagrams with extrinsic losses. Wavy lines denote the bare
         Coulomb interaction. Pluses and minuses indicate to which branch of the Keldysh
         contour [Fig.~\ref{fig:contour}(a)] corresponding points are belonging.}
      \label{fig:el}
\end{figure}

\begin{figure*}[t!]
	\centering
       \includegraphics[width=0.8\textwidth]{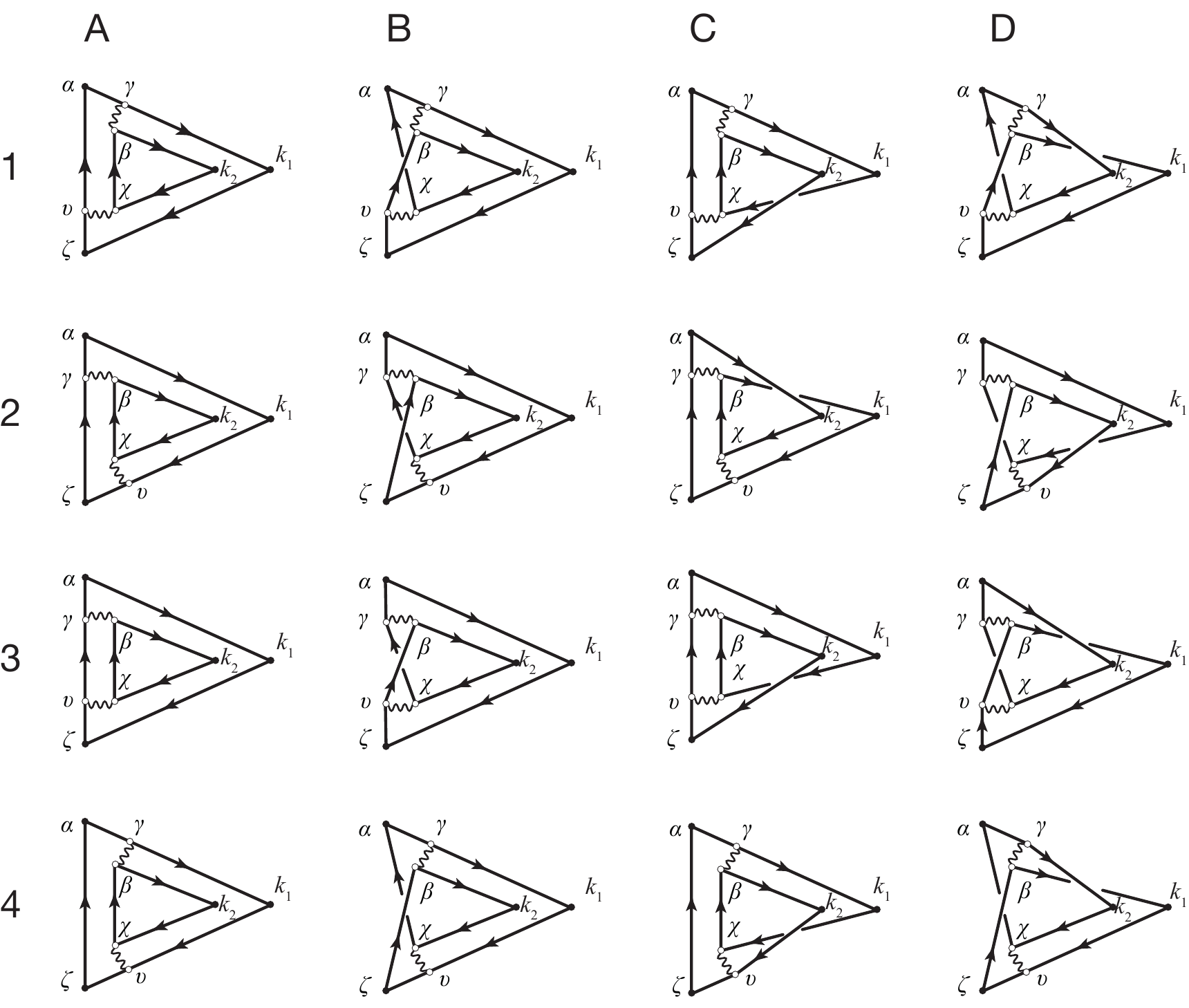}
       \caption[]{Leading order diagrams describing double photoemission. Notation 1A, 1B,
         etc., is used. Exchange of $\vec{k}_1$ and $\vec{k}_2$ generates topologically
         new ones. Corresponding columns will be denoted as A$^\prime$, B$^\prime$, and so
         on.
      \label{fig:diag_dpe_lord}}
\end{figure*}

There are several possibilities to introduce double photoemission.  One way is to regard
it as a single photoemission process acompanied by extrinsic energy losses whereby a
liberated electron interacts with the target system knocking out a secondary electron.  In
atomic, molecular and optical (AMO) physics
community~\cite{amusia_double_1968,amusia_atomic_1990} this is known as the \emph{knock
  out} mechanism. As an illustration consider second order diagrams at
Fig.~\ref{fig:el}. They are discussed in details by Caroli \emph{et al}. as examples of
extrinsic losses (cf. Fig.~9 of Ref.~\cite{caroli_inelastic_1973}).
%---------------------------
%---------------------------
There are no first order processes because bare Coulomb interaction is local in time and,
therefore, cannot connect points on different branches of Keldysh contour. In all three
cases there is a fermionic line flowing from ``$-$'' to ``$+$'' vertix, i.\,e., given by
the greater GF. This is a potential candidate for a secondary electron that can be
detected in the coincidence measurement. Intuitively corresponding DPE diagrams can be
drawn by cutting this line and adding an intermediate point with well defined momentum
$\vec{p}_2$. These and several more diagrams describing double photoemission can be
obtained applying the Wick theorem to the following
correlator~\cite{pavlyukh_single-_2015}:
%---------------------------
\begin{align}
Z^{(2)}(t,t^\prime)&=\langle\Psi_0|c_b^\dagger(t)c_a(t)c_{\vec{k}_1}^\dagger(0)c_{\vec{k}_2}^\dagger(0)\nn\\
&\quad\quad\quad\times c_{\vec{k}_2}(0)c_{\vec{k}_1}(0)c_c^\dagger(t^\prime)c_d(t^\prime)|\Psi_0\rangle.
\label{eq:Z2}
\end{align}
%---------------------------
The derivation is simpler if this correlator is written in the real space representation.
For completeness all leading order (second order in bare Coulomb interaction) diagrams are
shown on Fig.~\ref{fig:diag_dpe_lord}, where points labelled by $\alpha,\,\beta,\,\gamma$
denote integration over space, summation over spin, and integration over the time on
``$-$''-branch of the Keldysh contour. Similarly, $\chi,\,\upsilon,\,\zeta$ correspond to
the points on the ``$+$''-branch.  Specifically,
$\alpha\equiv(\vec{x}_\alpha,t_\alpha)\equiv(\vec{r}_\alpha,\sigma_\alpha,t_\alpha)$,
$\chi\equiv(\vec{x}_\chi,t_\chi)\equiv(\vec{r}_\chi,\sigma_\chi,t_\chi)$, etc.
%---------------------------
%---------------------------
Let us recall explicit form of the non-interacting greater GF:
\be
g_{\vec{x}_\alpha\vec{x}_\zeta}^>(t,t')=-i\sum_a\bar{f}(\epsilon_a)e^{-i\epsilon_a(t-t')}
\langle\vec{x}_\alpha|a\rangle\langle a|\vec{x}_\zeta\rangle,
\ee
%---------------------------
where $f(\epsilon)=1-\bar{f}(\epsilon)$ ist the zero temperature Fermi distribution
function and the sum runs over all one-particle states. For times
$t,\,t^\prime\in(-\infty,0]$ the product of two greater GFs can be expressed as:
%---------------------------
\begin{align}
  &\sum_{\vec{y}}g_{\vec{x}_\alpha\vec{y}}^>(t,0)g_{\vec{y}\vec{x}_\zeta}^>(0,t')\nn\\
  &=(-i)^2\sum_{ab}\bar{f}(\epsilon_a)\bar{f}(\epsilon_b)e^{-i\epsilon_at+i\epsilon_bt'}\sum_{\vec{y}}
\langle\vec{x}_\alpha|a\rangle\langle a|\vec{y}\rangle
\langle\vec{y}|b\rangle\langle b|\vec{x}_\zeta\rangle\nn\\
&\quad\quad\quad=(-i)^2\sum_{a}e^{-i\epsilon_a(t-t')}
\bar{f}(\epsilon_a)\langle\vec{x}_\alpha|a\rangle\langle a|\vec{x}_\zeta\rangle\nn\\
&\quad\quad\quad=-ig_{\vec{x}_\alpha\vec{x}_\zeta}^>(t,t'),
\end{align}
%---------------------------
where the completeness of the one-particle basis was used. The sum over $\vec{y}$ can be
done as integration over space coordinates and spin indices or in any other complete
basis. For instance we can perform the sum in the basis of single-particle states, i.\,e.
$\sum_{\vec{y}}\rightarrow\sum_i+\sum_{\vec k}$, where $i$ labels bound states and
$\vec{k}$ is the continuum state labeled by its momentum. Cutting a greater
non-interacting GF amounts to specification of an intermediate state,
e.\,g. $g_{\vec{x}_\alpha\vec{x}_\zeta}^>(t,t')\rightarrow
g_{\vec{x}_\alpha\vec{k}}^>(t,0)g_{\vec{k}\vec{x}_\zeta}^>(0,t')$. To the left hand side
we see a single GF that desribes the propagation of an electron. Its intermediate state is
not determined, whereas the product ot two GFs at the right hand side describes
observation of the intermediate state in the detector. This procedure can be used to
generate double photoemission diagrams starting from that of single photoemission
accompanied by energy losses.  Considering diagrams on Fig.~\ref{fig:el} we see that they
in particular transform into the diagrams 4A, 4A$^\prime$, and 4C, respectively, as
shown on Fig.~\ref{fig:diag_dpe_lord}. This figure contains a number of other
topologically distinct species which can be obtained by considering in addition to
extrinsic losses, the diagrams describing intrinsic losses, and interference of these two
mechanisms.
%---------------------------
\begin{figure}[b!]
	\centering
       \includegraphics[width=0.8\columnwidth]{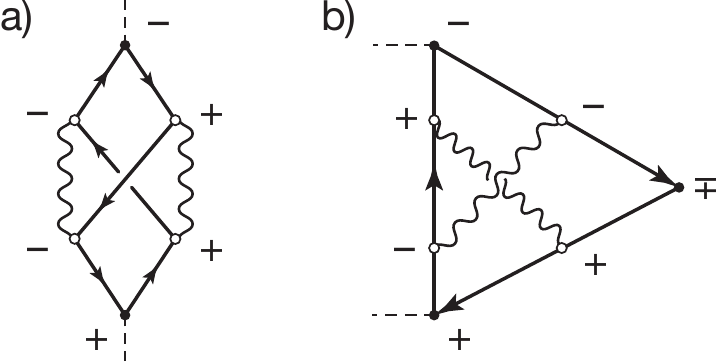}
       \caption[]{Application of cutting procedure to the lesser density-density response
         function (left) yields a SPE diagram (right) which, in turn, leads to the 4C-type
         diagram shown on Fig.~\ref{fig:diag_dpe_lord}.}
      \label{fig:il}
\end{figure}
%---------------------------
The outlined procedure can be generalized to obtain SPE diagrams from that of the
density-density response function $\chi^{<}(t,t')$, DPE from SPE, and so on:
\begin{subequations}
\begin{align}
Z^{(1)}_{\vec{k}}(t,t')&=\int d(\alpha\zeta) \frac{\delta \chi^<(t,t')}
{\delta g^>_{\vec{x}_\alpha\vec{x}_\zeta}(\tau_\alpha,\tau_\zeta)}
g_{\vec{x}_\alpha\vec{k}}^>(\tau_\alpha,0)g_{\vec{k}\vec{x}_\zeta}^>(0,\tau_\zeta),\\
Z^{(2)}_{\vec{k}\vec{k}^\prime}(t,t')&=\int d(\gamma\upsilon)\frac{\delta Z^{(1)}_{\vec{k}}(t,t')}
{\delta g^>_{\vec{x}_\gamma\vec{x}_\upsilon}(\tau_\gamma,\tau_\upsilon)}
g_{\vec{x}_\alpha\vec{k}^\prime}^>(\tau_\gamma,0)g_{\vec{k}^\prime\vec{x}_\zeta}^>(0,\tau_\upsilon).
\end{align}
\end{subequations}
In this way the 4C-type diagram is a descendant of the SPE diagram in Fig.~\ref{fig:il}(b)
which, in turn, traces back to the density-density response function in
Fig.~\ref{fig:il}(a).

It is more then just graphs operations that is illustrated here. In
ref.~\cite{almbladh_photoemission_2006} and in a sequence of
works~\cite{stefanucci_diagrammatic_2014,uimonen_diagrammatic_2015} it has been shown that
cutting procedure, which splits diagrams representing the lesser or the greater response
function into \emph{half-diagrams}, can be used to construct approximations with the
positive definiteness property. These are the approximations for response functions
possessing positive spectral functions in the whole frequency range. Explicit construction
outlined there shows that in this case the spectral function can be written in the FGR
form, that is, as a sum of terms containing squared modulus of the matrix element of a
certain scattering process multiplied with corresponding $\delta$-function assuring
overall energy conservation. The matrix element then is expressed in terms of GFs on only
one branch of the Keldysh contour, i.\,e. all its vertices are labeled with either ``$+$''
or with ``$-$''.

\begin{figure}[]
\centering
\includegraphics[width=\columnwidth]{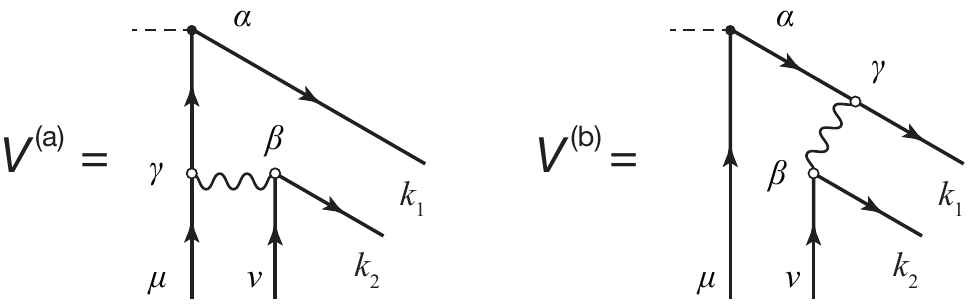}
\caption[]{Half-diagrams corresponding to Fig.~\ref{fig:diag_dpe_lord}. Same labeling is
  preserved. All vertices have time arguments on ``$-$''-branch of the Keldysh contour.
  $V^{(a)}$ has interaction line connected to the lesser propagator for primary electron,
  the scattering reflects intrinsic losses in the ground state. In AMO physics this
  process is also known as \emph{shake off} mechanism. In contrast, $V^{(b)}$ has
  interaction line connected to the greater propagator for primary electron, the
  scattering reflects extrinsic losses in excited state. In AMO physics this process is
  also known as \emph{knock out} mechanism. There two more topologically equivalent
  structures with exchanged $\vec{k}_1$ and $\vec{k}_2$ indices. They will be denoted as
  $\bar{V}^{(a)}$ and $\bar{V}^{(b)}$. \label{fig:half}}
\end{figure}

\begin{figure}[b]
\centering
\includegraphics[width=\columnwidth]{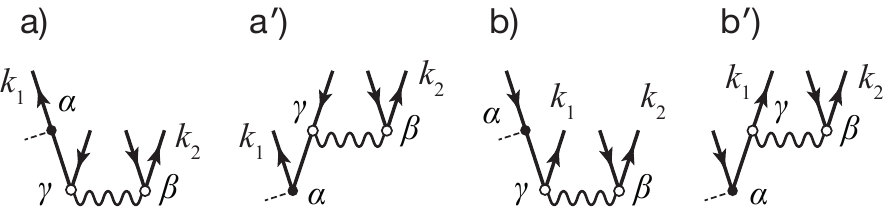}
\caption[]{Goldstone diagrams corresponding to the matrix elements of double photoemission
  in Fig.~\ref{fig:half}. For NEGF diagrams $+$ and $-$ assigned to the vertices partly
  determine the time ordering; here, the direction of arrows specifies whether particle
  (arrow up) or hole (arrow down) propagators should be used. Therefore, each scattering
  matrix element shown in Fig.~\ref{fig:half} is represented by two diagrams: $V^{(a)}$ by
  (a) and (a'), $V^{(b)}$ by (b) and (b'). This reflect the fact that in addition to
  dangling lines which are described by either particle or hole propagators, $V^{(a)}$ and
  $V^{(b)}$ contain one internal $g^{--}$ line. It is given by \emph{time-ordered}
  propagator comprising both the particle and the hole terms.
\label{fig:gold}}
\end{figure}

For scattering problems as considered here the Fermi Golden rule approach is particularly
useful. It allows to describe arbitrary scattering processes in leading order without
referring to complicated mathematical apparatus of Feynman diagrams. It is also the only
choice for more complicated processes if perturbative expressions cannot readily be
generated. Therefore, it is interesting to identify the ingredients of diagrams on
Fig.~\ref{fig:diag_dpe_lord}, i.\,e., the half-diagrams yielding corresponding expressions
in the FGR form. Independently whether the second-order density-density response diagrams
(many of them are listed by Uimonen \emph{et al}.~\cite{uimonen_diagrammatic_2015}) are
used as a starting point, or the single photoemission diagrams listed for instance by
Caroli \emph{et al.}~\cite{caroli_inelastic_1973}, or the one shown here on
Fig.~\ref{fig:diag_dpe_lord}, the result will be topologically identical. The difference
is in the treatment of dangling lines: in the former case their quantum numbers are summed
over, whereas in SPE one dangling line is observed in the detector and therefore has a
fixed index. For emission of $n$-particles by one photon there should be $n$ dangling
lines corresponding to momenta registered (in coincidence) in the detectors.  Rather
surprisingly, the result of such analysis (Fig.~\ref{fig:half}) is quite simple and can be
recovered from early works of Amusia and Kazachkov~\cite{amusia_double_1968}, they can
also be identified as diagrams in Fig.~7(b) in Uimonen \emph{et
  al.}~\cite{uimonen_diagrammatic_2015}.
%---------------------------
%---------------------------
In order to make explicit parallels with works of AMO physics community consider Goldstone
diagrams (see Chang and McDowell~\cite{chang_photo-ionization_1968} for a short summary of
this technique in application to atomic photoionization and L'Huillier \emph{et
  al}.~\cite{lhuillier_many-electron_1986} for more complicated multiphoton examples) that
are commonly used in scattering theory. To make this work self-contained corresponding
lowest order DPE Goldstone diagrams are shown in Fig.~\ref{fig:gold} (cf. Fig.~8.1 in
ref.~\cite{amusia_atomic_1990}).

%---------------------------
%---------------------------
In fact, Goldstone diagrams carry more information than standard NEGF ones: the direction
of arrow on each fermionic line specifies whether particle or hole propagator should be
used. For the latter, ``$+$'' and ``$-$'' assigned to the vertices only specify the branch
of Keldysh contour they belong to; however, all possible mutual ordering withing the
branch have to be considered. That is why there are twice as many diagrams in
Fig.~\ref{fig:gold} as compared to Fig.~\ref{fig:half}: the two possibilities correspond
to the two terms in the time-ordered GF of internal line.

Let us verify that all triangular DPE diagrams at Fig.~\ref{fig:diag_dpe_lord} can be
obtained by gluing half-diagrams in Fig.~\ref{fig:half}. The simplest identities would be
%---------------------------
\bea
Z^{A}&=&\sum_{\mu\nu}|V^{(a)}+V^{(b)}|_{\mu\nu}^2,\nn\\
Z^{A'}&=&\sum_{\mu\nu}|\bar{V}^{(a)}+\bar{V}^{(b)}|_{\mu\nu}^2,\nn
\eea
%---------------------------
where $Z^{A}$ denotes the sum of contributions from column ``A'' of
Fig.~\ref{fig:diag_dpe_lord}, the indices $\mu$ and $\nu$ label dangling lines, and
momentum dependence, i.\,e. $\vec{k}_{1,2}$, is not shown explicitly. This is not the full
set because, as explained in
refs.~\cite{stefanucci_diagrammatic_2014,uimonen_diagrammatic_2015} permutations of
dangling lines must also be considered:
\[
Z^{B}=\sum_{\mu\nu}(V^{(a)}+V^{(b)})_{\mu\nu}(V^{(a)}+V^{(b)})^*_{\nu\mu}.
\]
Collecting all contributions the full set at Fig.~\ref{fig:diag_dpe_lord},
i.\,e. $Z^{(2)}=Z^{A}+Z^{A'}+Z^{B}+Z^{B'}+Z^{C}+Z^{C'}+Z^{D}+Z^{D'}$ can be obtained as
follows:
%---------------------------
\be
Z^{(2)}=\sum_{\mu\nu}(V+\bar{V})_{\mu\nu}
(V+V^\mathrm{T}+\bar{V}+\bar{V}^\mathrm{T})^*_{\mu\nu},
\ee
%---------------------------
where the sum of diagrams in Fig.~\ref{fig:half} is denoted by $V=V^{(a)}+V^{(b)}$. There
are 32 terms in this expression. The construction guarantees the positivity of
corresponding two-particle current.

\section{Generalizations}
This construction can be generalized to the process of emission of arbitrary number of
particles. It is clear that higher order diagrams need to be considered. Already an
example of classical system of particles with two-body hard sphere interactions shows that
emission of $n$ particles requires at least $n-1$ collisions. Quite generally, in order to
achieve $n$-particle emission Goldstone diagrams must contain at least $n-1$ interaction
lines, the scattering cross-section is being proportional to $v^{2(n-1)}$ as shown in
ref.~\cite{pavlyukh_single-_2015}. However, what can be said about topological structure
of these diagrams?

One possibility to characterize diagrammatic complexity is to enumerate possible
topological structures. For equilibrium many-body perturbation theory the problem of
counting diagrams was initiated in the work of Molinari~\cite{molinari_hedins_2005} who
suggested to use the Hedin's system of functional
equations~\cite{hedin_new_1965,strinati_application_1988} that relates the electron GF,
the proper self-energy $\Sigma$, vertex function $\Gamma$, polarization $P$, and the
effective potential $W$ as the generating functions for corresponing diagram counting. The
transformation from integral-functional equations to much simpler differential equations
for generating functions is done by collapsing all space, spin and time coordinates into
one point, i.\,e., considering zero-dimensional case. These equations have been
subsequently solved analytically~\cite{pavlyukh_analytic_2007}. Besides pure mathematical
interest, the solution can be used as a basis for method
development~\cite{lani_approximations_2012,pavlyukh_taming_2016} and to get insight into
the convergence of approximate theories~\cite{berger_solution_2014,stan_unphysical_2015}.

We have seen above that arbitrary photoemission process can be described using the lesser
density-density response function ($\chi^{<}$) as a starting point. Corresponding
time-ordered irreducible response ($P^{--}$) is the polarization in Hedin's system of
equations~\cite{strinati_application_1988}. Thus, it seems natural to use the
nonequilibrium extension of these equations~\cite{harbola_nonequilibrium_2006,
  ness_gw_2011} in order to enumerate \emph{decorated} Feynman diagrams (with ``$+$'' and
``$-$'' assigned to the vertices). Below, these equations are explicitly written
discarding the time and space dependence of all quantities (in zero dimensions), but
preserving the labeling of vertices:
%========================
\begin{subequations}
\label{eq:hedin}
\begin{align}
\Sigma^{\eta\bar{\eta}}&=i\sum_{\eta_3\eta_4}\bar{\eta}_3\bar{\eta}_4
W^{\eta\eta_3}G^{\eta\eta_4}\Gamma^{\eta_4\bar{\eta},\eta_3},\\ 
P^{\eta\bar{\eta}}&=-i\sum_{\eta_3\eta_4}\bar{\eta}_3\bar{\eta}_4
G^{\eta\eta_3}G^{\eta_4\eta}\Gamma^{\eta_3\eta_4,\bar{\eta}},\label{eq:P}\\ 
\Gamma^{\eta_1\eta_2,\eta_3}&=\delta_{\eta_1\eta_2}\delta_{\eta_2\eta_3}+g^{\eta_4\eta_3}g^{\eta_3\eta_5}
\frac{\delta\Sigma^{\eta_1\eta_2}}{\delta g^{\eta_4\eta_5}},\label{eq:GM}\\ 
G^{\eta\bar\eta}&=g^{\eta\bar\eta}+g^{\eta\bar\eta}\Sigma^{\bar\eta\bar\eta}G^{\bar\eta\bar\eta}
-g^{\eta\eta}\Sigma^{\eta\bar\eta}G^{\bar\eta\bar\eta}
+g^{\eta\eta}\Sigma^{\eta\eta}G^{\eta\bar\eta},\\ 
W^{\eta\bar\eta}&=
vP^{\eta\eta}W^{\eta\bar\eta}-vP^{\eta\bar\eta}W^{\bar\eta\bar\eta} .
\end{align}
\end{subequations}
%========================
Here $\eta=\pm1$ denotes branches of Keldysh contour, $\bar\eta=-\eta$. In equations for
the screened interaction some terms are not present because the bare interaction is
instantaneous, i.\,e. $v^{\eta\bar\eta}=0$. One more assumption was incorporated: in
Dyson's equations for GFs it was explicitly used that in equilibrium (or in steady state)
and at zero temperature the following convolutions vanish $\int\!
d\bar{t}\,X^{\eta\bar\eta}(t,\bar{t})Y^{\bar\eta\eta}(\bar{t},t')=0$ (see discussion by
Ness \emph{et al.}~\cite{ness_gw_2011}). Under this assumption equations for time-ordered
and anti-time-ordered functions are reduced to the usual form. From the structure of
equations~\eqref{eq:hedin} the vertices are necessarily decorated such that no isolated
islands of pluses or minuses
appear~\cite{stefanucci_diagrammatic_2014,uimonen_diagrammatic_2015}.

%---------------------------
\begin{figure}[]
\centering
\includegraphics[width=\columnwidth]{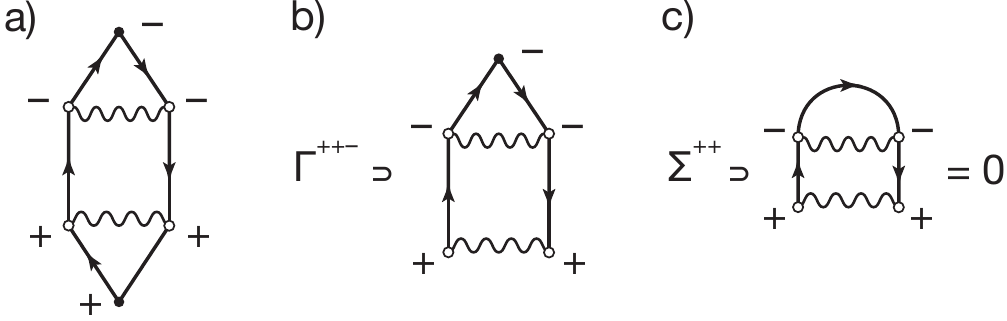}
\caption[]{(a) Example of polarization that cannot be obtained from Eq.~\eqref{eq:P}.
  Corresponding vertex function (b) can be constructed by applying Eq.~\eqref{eq:GM} to
  self-energy (c).
\label{fig:anom}}
\end{figure}
%---------------------------

Let us now skip $\pm i$ prefactors and signs (not essential for the diagram enumeration),
introduce the prefactor $\ell$ into eq.~\eqref{eq:P} in order to keep count of fermionic
loops and write the power series solution for $P^<$:
\begin{align*}
  P^{-+}&=3\ell v{g^{-+}} {g^{+-}} \big\{[g^{--}]^2+[g^{++}]^2\big\}\\
&+v^2 {g^{-+}} {g^{+-}} \Big\{(5 \ell^2+15 \ell)\big([g^{--}]^4 +[g^{++}]^4\big)\\
&+6 \ell[g^{--}]^2 [g^{++}]^2
  +(8 \ell ^2+8 \ell) g^{--}g^{-+}g^{+-}g^{++}\Big\}\\
  &+\mathcal{O}(v^3).
\end{align*}
The $(8 \ell ^2+8 \ell)$ term numbers all diagrams in Fig.~\ref{fig:diag_dpe_lord},
however, three other terms of the second order (they contain a single greater propagator
and therefore contribute only to SPE, but not to DPE) are missing [$3\ell v^2 {g^{-+}}
  {g^{+-}}[g^{--}]^2 [g^{++}]^2$; one such example is shown in
  Fig.~\ref{fig:anom}(a)]. This can be traced back [Fig.~\ref{fig:anom}(c)] to the
self-energy $\Sigma^{\eta\eta}$ in the vertex equation~\eqref{eq:GM}. Per construction
(recall that we impose $X^{\eta\bar\eta}Y^{\bar\eta\eta}=0$) it contains only
$g^{\eta\eta}$ lines and is therefore insufficient to obtain the vertex function shown in
Fig.~\ref{fig:anom}(b).  Thus, a mere extension of Hedin's equations to nonequilibrium
case does not allow to enumerate photoemission diagrams. Spurious diagrams appear if
restriction $X^{\eta\bar\eta}Y^{\bar\eta\eta}=0$ is lifted, otherwise valid diagrams are
missing. There is, actually, a logical explanation to this fact: due to asymmetric
position of the vertex function the Hedin's expressions for the self-energy and
polarization do not respect the FGR form pertinent to exact $\Sigma^{\eta\bar{\eta}}$ and
$P^{\eta\bar{\eta}}$.
\section{Conclussions}
In this work the NEGF diagrams for single- and multi-electron emission were treated
consistently by using the lesser density-density response ($\chi^<$) as a common
generating function. In this way $n$-particle emission is described by diagrams containing
$n$ lesser electron propagators with a well defined momentum.  Alternatively, the
photoemission cross-section can be recasted in the Fermi Golden rule form. It is common to
write corresponding matrix elements in terms of Goldstone diagrams. Gluing such diagrams
relating to the time-propagation in forward, backward directions yields the NEGF
construction. This correspondence was illustrated by considering all possible double
photoemisison processes to the leading order in Coulomb interaction. Equivalence between
the shake off and knock out mechanisms known in AMO physics and intrisic and extrinsic
losses in condensed matter physics is demonstrated.

Since $\chi^<$ is a central quantity of this study it is desirable to have a set of
functional relations that would allow to recursively generate its perturbative expansion
in a manner similar to Hedin's equations for time-ordered quantities. An extension of
these equations to nonequilibrium case generates a larger set of diagrams, some of them
being zero at zero temperature and in a steady state. Because Hedin's equations for
self-energy and polarization are not compatible with FGR these spurious diagrams cannot be
easily eliminated.

\section*{Acknowledgments} It is a pleasure to thank the organisers Claudio Verdozzi,
Andreas Wacker and Carl-Olof Almbladh for an excellent workshop on \emph{Progress in
  Nonequilibrium Green's Functions} (PNGF6).  This work is supported by the DFG grants
No. SFB762 and No. PA 1698/1-1.
%\bibliography{MyLibrary}
%

\end{document}